\documentclass[superscriptaddress,showpacs,aps,twocolumn,prb,floatfix,nobibnotes]{revtex4-2}
\usepackage{amssymb}
\usepackage{amsmath}
\usepackage{mathtools}
\usepackage{graphicx}
\usepackage{bm}
\usepackage{fontenc}
\usepackage{color}
\usepackage{ulem} 
\usepackage{varioref}
\usepackage[hidelinks]{hyperref}
\begin{document}
\title{Low-energy states of a Su-Schrieffer-Heeger chain
connected perpendicularly to a metallic lead}

\author{A. A. Aligia}
\email{aaligia@gmail.com}
\affiliation{Instituto de Nanociencia y Nanotecnolog\'{\i}a
CNEA-CONICET, GAIDI,
Centro At\'{o}mico Bariloche, 8400 Bariloche, Argentina}

\begin{abstract}
Recently, Su-Schrieffer-Heeger chains (SSHCs) coupled to metallic leads have attracted attention. In particular Mart\'{\i}nez {\it et al.}
[Phys. Rev. Res. {\bf 8}, 043118 (2026)] studied a
topological SSHC connected at one of its ends perpendicularly to a metallic lead modeled by a tight-binding chain.
Here, we extend their results by incorporating the finite-size splitting of the end modes of the isolated SSHC, which are determined analytically with high accuracy.
We also derive simple analytical expressions, valid in the low-energy regime, for the zero-temperature conductance through the lead and the relevant spectral densities. We find that, when the chemical potential coincides with that of the half-filled SSHC, the conductance reaches its maximum value, equal to that of the lead in the absence of the SSHC.
By contrast, the conductance vanishes when the chemical potential is shifted by an amount equal to half the end-mode splitting.
\end{abstract}

\maketitle

\section{Introduction}
\label{intro}
%\noindent

In the last two decades, topological materials have aroused considerable interest \cite{Ryu10,Qi11,Ando13,Asb16}.
Several experimental studies have realized extensions of the Su-Schrieffer-Heeger chain (SSHC) \cite{Li24,Jean17,Gagel24,Yu25}. In its simplest form, the SSHC introduced to describe polyacetylene \cite{Su79}, consists of a dimerized chain of spinless fermions with alternating nearest-neighbor couplings.
It supports two distinct topological phases, characterized by a
$\mathbb{Z}_2$ topological invariant. In particular, for an appropriate choice of unit cell, the Zak phase takes the values $0$ and $\pi$ in the trivial and topological phases, respectively \cite{Asb16}, corresponding to winding numbers of 0 and 1. In the topological phase of a finite chain, the absolute values of the first and last hoppings are smaller than those of the other hoppings. In this case, if the chain is infinite, a zero-energy mode appears at each end of the chain.

In recent years, numerous extended SSHCs have been investigated \cite{Li24,Jean17,Gagel24,Yu25,Dias22,Aligia23,Ghosh23,Malakar23,Ghuneim24,Joshi25,Chang25,Ali25,He26,Grizzi26,Zwil26,Gri2}, some of which support integer winding numbers beyond 0 and 1  \cite{Li24,Dias22,Ghosh23,Malakar23,Joshi25,Chang25,Grizzi26}.
In addition, SSHCs (some including interactions) coupled to other systems, such as scanning tunnelling microscope (STM) tips \cite{Zhao24,Ali25} or metallic leads \cite{Wagner23,Bisson24,Mart26}, have also been investigated recently.

In particular, Mart\'{\i}nez et al. \cite{Mart26} studied a topological SSHC coupled at one of its ends to the midpoint of a metallic lead modeled by a tight-binding chain. For a sufficiently long SSHC, the end state located at the opposite end of the chain, far from the metallic lead, becomes a topologically protected bound state in the continuum at zero energy, which is of particular interest. Their work extended a previous approach in which the states of an infinitely long SSHC were considered \cite{Bisson24}. However, in a finite chain, the two end states hybridize, resulting in a finite energy splitting away from zero. This finite-size effect modifies the low-energy spectral densities and the conductance through the metallic lead.
The goal of the present work is to describe these modifications using a simple yet highly accurate analytical approach valid in the low-energy regime.

The paper is organized as follows. In Section \ref{model}, we introduce the model and the approximation used. Section \ref{res} contains the results and Section \ref{sum} summarizes and discusses our results.

\section{Model and approximation}

\label{model}

The starting model is the same as that considered by
Mart\'{\i}nez \textit{et al.} \cite{Mart26}.  The Hamiltonian is

\begin{equation}
H=H_{\text{SSHC}}+H_{\text{ML}}+H_{\text{C}},  \label{ham}
\end{equation}
where the first term corresponds to an isolated SSHC of $N$ unit cells (2$N$
sites)

\begin{equation}
H_{\text{SSHC}}=-v\sum_{i=1}^{N}\left( a_{i}^{\dagger }b_{i}+\text{H.c.} \right)
-w\sum_{i=1}^{N-1}\left( b_{i}^{\dagger }a_{i+1}+\text{H.c.}\right) ,
\label{hssh}
\end{equation}
the second term corresponds to the metallic lead

\begin{equation}
H_{\text{ML}}=-J\sum_{j}\left( c_{j}^{\dagger }c_{j+1}+\text{H.c.}\right) ,
\label{hml}
\end{equation}
and the last term is the coupling between them

\begin{equation}
H_{\text{C}}=-t_{c}c_{0}^{\dagger }a_{1}+\text{H.c.}  \label{hc}
\end{equation}
With an appropriate choice of phases of all the operators, one can assume $v,w,J,t_{c}>0$.

In the limit $N\rightarrow \infty $, in the topological phase that takes
place when $v<w$, $H_{\text{SSHC}}$ has two zero-energy modes, one at the
left end of the chain $l^{\dagger }|0\rangle $ and another at right one $r^{\dagger }|0\rangle $, where \cite{Bisson24,Ali25}

\begin{eqnarray}
l^{\dagger } &=&p\sum\limits_{j=1}^{N}z^{j-1}a_{j}^{\dagger },  \label{op_l}
\\
r^{\dagger } &=&p\sum\limits_{j=1}^{N}z^{N-j}b_{j}^{\dagger },  \label{op_r}
\\
z &=&-v/w,\text{ }p=\sqrt{\frac{1-z^{2}}{1-z^{2N}}}.  \label{z}
\end{eqnarray}
For finite chains and absolute value of the energies below half of the gap $E_{g}=w-v$ one has \cite{Ali25}

\begin{eqnarray}
H_{\text{SSHC}} &\simeq &H_{\text{SSHC}}^{\text{LE}}=-t_{lr}l^{\dagger }r+%
\text{H.c.,}  \notag \\
t_{lr} &=&\frac{1-z^{2}}{1-z^{L}}z^{N-1}v,  \label{app}
\end{eqnarray}
where the superscript LE stands for low energy. This is the only
approximation made in this work. The relative error introduced by the
states at energies outside the gap ($E<-E_{g}$ or $E>E_{g}$) is
of order $z^{N+2}/(1+z)$ \cite{Ali25} and decays
exponentially with increasing $N$.
Comparison with numerical data for the eigenergies near zero is excellent
(Fig. 6 of Ref. \cite{Ali25}). An extension of this approach to the SSHC
including other hoppings also shows an excellent agreement with numerical
data for the low-energy eigenstates  \cite{Grizzi26}.

Projecting Eq. (\ref{hc}) in the low-energy sector one obtains

\begin{eqnarray}
H_{\text{C}} &\simeq &H_{\text{C}}^{\text{LE}}=-t_{cl}c_{0}^{\dagger }l+\text{H.c,}  \notag \\
t_{cl} &=&pt_{c}.  \label{hcle}
\end{eqnarray}

\section{Results}

\label{res}

For convenience, we write the low-energy part of the Hamiltonian as $H^{\text{LE}}=H^{0}+V$, where $H^{0}=H_{\text{SSHC}}^{\text{LE}}+H_{\text{ML}}$
and $V=H_{\text{C}}^{\text{LE}}$. We define the Green function operators $\hat{G}=(\omega -H^{\text{LE}})^{-1}$ and \ $\hat{G}^{0}=(\omega -H^{0})^{-1}$.
We are particularly interested in the Green function at site 0 of the
metallic lead, $\hat{G}_{00}(\omega )=\langle \langle c_{0};c_{0}^{\dagger
}\rangle \rangle _{\omega }$, which as shown below, determines the
conductance through the metallic lead. The Green functions for $V=0$ are the
matrix elements of  $\hat{G}^{0}$ and are known. The Green functions of the
complete system can be found from the matrix elements of the equation

\begin{equation}
\hat{G}=\hat{G}^{0}+\hat{G}^{0}V\hat{G},  \label{gg0}
\end{equation}
derived from the definitions of $\hat{G}$ and $\hat{G}^{0}$.

The conductance $G$ through the metallic lead can be expressed in terms of the spectral density  $\rho _{0}(\omega )=(-1/\pi )\text{Im}\hat{G}_{00}(\omega )$, following the Meir-Wingreen formalism \cite{MW}. The resulting expression is analogous to that for transport through a quantum dot embedded between metallic leads \cite{Ali02}, and is given by

\begin{eqnarray}
G &=&\pi G_{0}\int d\omega \left( -\frac{f(\omega )}{d\omega }\right)
\Gamma (\omega )\rho _{0}(\omega ),  \notag \\
G_{0} &=&\frac{e^{2}}{h},\text{ }\Gamma (\omega )=\sqrt{4J^{2}-\omega ^{2}},
\label{gt}
\end{eqnarray}
where $f(\omega )$ is the Fermi function.

At zero temperature Eq. (\ref{gt}) reduces to

\begin{equation}
G=\pi G_{0}\sqrt{4J^{2}-\mu ^{2}}\rho _{0}(\mu ),  \label{cond}
\end{equation}
where $\mu $ is the chemical potential.

Taking the matrix element 00 of Eq. (\ref{gg0}) one obtains

\begin{equation}
\hat{G}_{00}=\hat{G}_{00}^{0}-\hat{G}_{00}^{0}t_{cl}\hat{G}_{l0},
\label{g00}
\end{equation}
and similarly from the element $l0$

\begin{equation}
\hat{G}_{l0}=-\hat{G}_{ll}^{0}t_{cl}\hat{G}_{00}.  \label{gl0}
\end{equation}
Replacing this result in the previous equation one obtains

\begin{equation}
\hat{G}_{00}=\hat{G}_{00}^{0}+t_{cl}^{2}\hat{G}_{00}^{0}\hat{G}_{ll}^{0}\hat{G}_{00},  \label{g00p}
\end{equation}
which allows to express $\hat{G}_{00}$ in terms of the Green functions of $H_{0}$. Using

\begin{equation}
\hat{G}_{00}^{0}=\frac{-i}{\sqrt{4J^{2}-\omega ^{2}}},\text{ }\hat{G}_{ll}^{0}=\frac{\omega }{\omega ^{2}-t_{lr}^{2}},  \label{g0v}
\end{equation}
where the first equation is valid for $|\omega |<J$, one obtains

\begin{eqnarray}
\hat{G}_{00}(\omega ) &=&\frac{-i}{\sqrt{4J^{2}-\omega ^{2}}+iC(\omega )},
\notag \\
C(\omega ) &=&\frac{t_{cl}^{2}\omega }{\omega ^{2}-t_{lr}^{2}}.  \label{g00u}
\end{eqnarray}
There are in addition two poles of  $\hat{G}_{00}(\omega )$ at
$\omega =\pm \left[ 2J^{2}+\sqrt{4J^{4}+t_{cl}^{4}}\right] ^{1/2}$.

Replacing in Eq. (\ref{cond}) one obtains for the conductance at zero
temperature

\begin{equation}
G(\mu )=\frac{4J^{2}-\mu ^{2}}{4J^{2}-\mu ^{2}+C^{2}(\mu )}.  \label{cond0}
\end{equation}

\begin{figure}[htb]
\centering
\includegraphics[width=0.8\columnwidth]{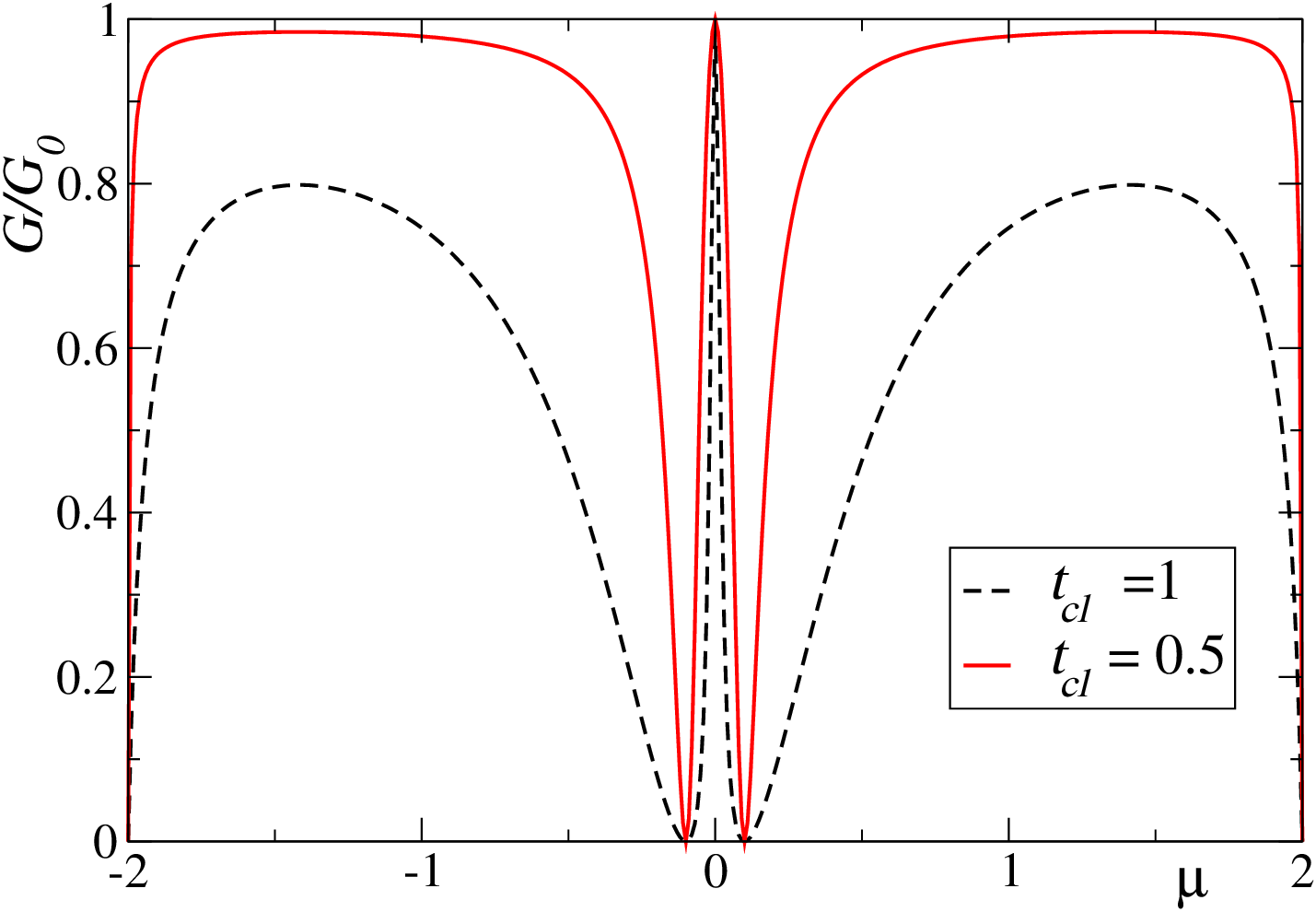}\\[0.8cm]
\includegraphics[width=0.8\columnwidth]{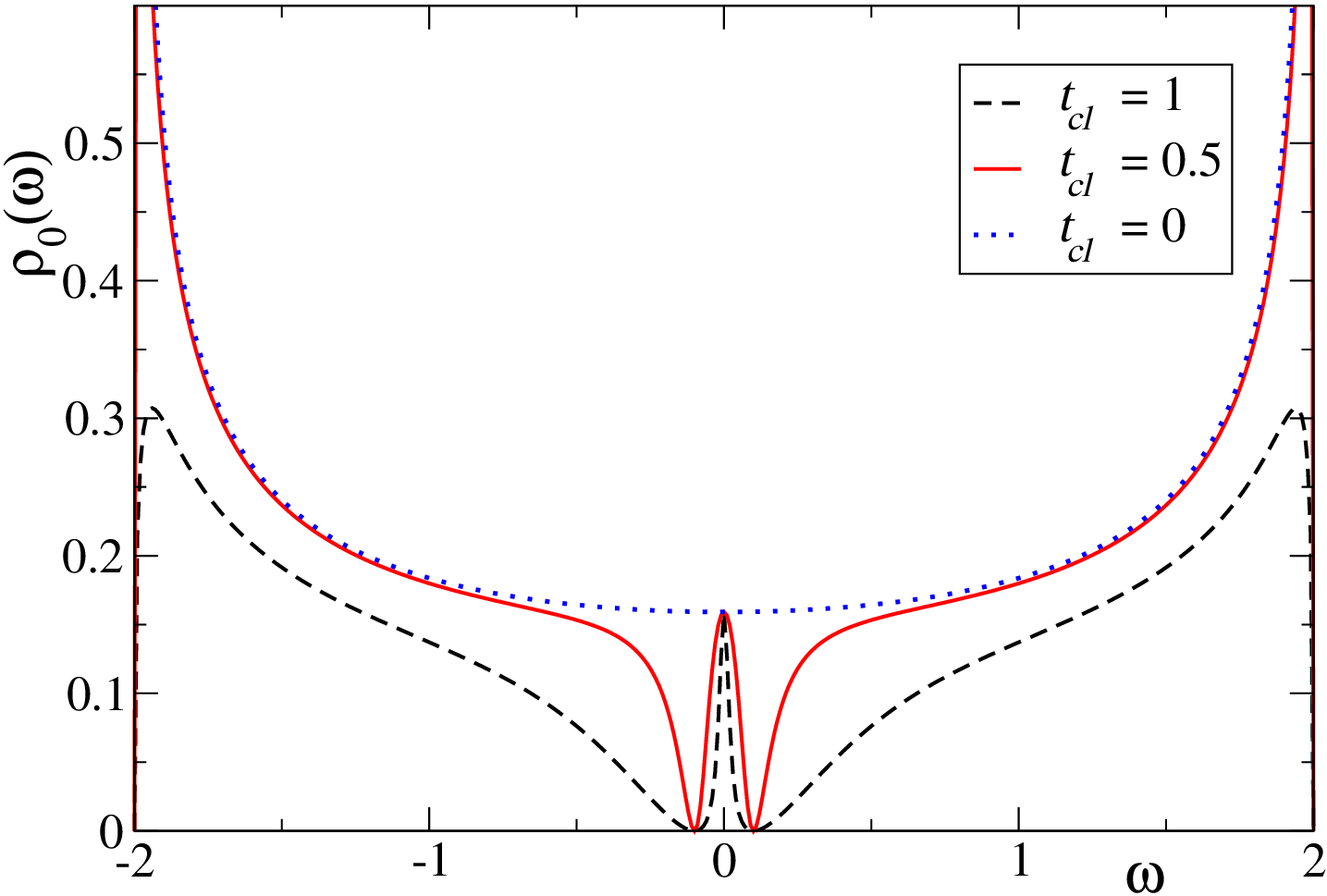}
\caption{(Color online) Top: zero-temperature conductance as a function of chemical potential. Bottom: spectral density
of the site of the metallic lead connected to the SSHC as
a function of energy.
Parameters are $J=1$, $t_{lr}=0.1$ and several values of $t_{cl}$.}
\label{gro0}
\end{figure}

In Fig. \ref{gro0}, we show this function and $\rho_0(\omega)$ for selected parameter values. Note that their dependence on the original SSHC parameters, $v$ and $w$, is only indirect, through $t_{lr}$ [see Eq. (\ref{app})] and $t_{cl}$ [see Eqs. (\ref{z}) and (\ref{hcle})]. We recall that our results are valid for $|\mu|,|\omega|<E_g$. Nevertheless, the metallic-lead hopping $J$ can be much smaller than the SSHC gap $2E_g$.
The conductance reaches the
quantum of conductance $G_0$ for $\mu=0$ and vanishes
for $\mu = \pm t_{lr}$. For other values of the chemical potential
the conductance is smaller but near to $G_0$, particularly
for small values of $t_{cl}$. These properties might be used
to detect experimentally the splitting $2t_{lr}$ of the
low-energy modes.

Similarly, the spectral density at site 0 is reduced but remains qualitatively similar to that of a disconnected metallic lead ($t_{cl}=0$), except at
$\omega=\pm t_{lr}$, where it vanishes, and in their vicinity. The reduction of the spectral weight within the interval $-2J<\omega<2J$, relative to the disconnected-lead case, is compensated by the appearance of two delta-function peaks at $\omega=\pm\left[2J^2+\sqrt{4J^4+t_{cl}^4}\right]^{1/2}$.

Proceeding in a similar way as before, one obtains

\begin{equation}
\hat{G}_{ll}(\omega )=\hat{G}_{ll}^{0}(\omega )\left( 1-\frac{it_{cl}^{2} \hat{G}_{ll}^{0}(\omega )}{it_{cl}^{2}\hat{G}_{ll}^{0}(\omega )+\sqrt{4J^{2}-\omega ^{2}}}\right) ,  \label{gll}
\end{equation}
where $\hat{G}_{ll}^{0}(\omega )$ is given in Eqs. (\ref{g0v}). From $\rho
_{l}(\omega )=(-1/\pi )\text{Im}\hat{G}_{ll}(\omega )$ one obtains the
density of states of the left mode in presence of the metallic lead

\begin{equation}
\rho _{l}(\omega )=\frac{1}{\pi }\frac{t_{cl}^{2}\omega ^{2}\sqrt{4J^{2}-\omega ^{2}}}{t_{cl}^{4}\omega ^{2}+(4J^{2}-\omega
^{2})(t_{cl}^{2}-\omega ^{2})^{2}}.  \label{roll}
\end{equation}
Finally $\hat{G}_{rr}(\omega )=\langle \langle r;r^{\dagger }\rangle \rangle
_{\omega }$ can \ be related to $\hat{G}_{ll}(\omega )=\langle \langle
l;l^{\dagger }\rangle \rangle _{\omega }$ using appropriate equations of
motion

\begin{eqnarray}
\omega \langle \langle r;r^{\dagger }\rangle \rangle _{\omega }
&=&1-t_{lr}\langle \langle l;r^{\dagger }\rangle \rangle _{\omega },  \notag
\\
\omega \langle \langle l;r^{\dagger }\rangle \rangle _{\omega }
&=&-t_{lr}\langle \langle l;l^{\dagger }\rangle \rangle _{\omega }.
\label{motion}
\end{eqnarray}
From here

\begin{eqnarray}
\hat{G}_{rr}(\omega ) &=&\frac{\omega +t_{lr}^{2}}{\omega ^{2}}\hat{G}_{ll}(\omega ),  \notag \\
\rho _{r}(\omega ) &=&\frac{t_{lr}^{2}}{\omega ^{2}}\rho _{l}(\omega ).
\label{ror}
\end{eqnarray}

\begin{figure}[htb]
\centering
\includegraphics[width=0.8\columnwidth]{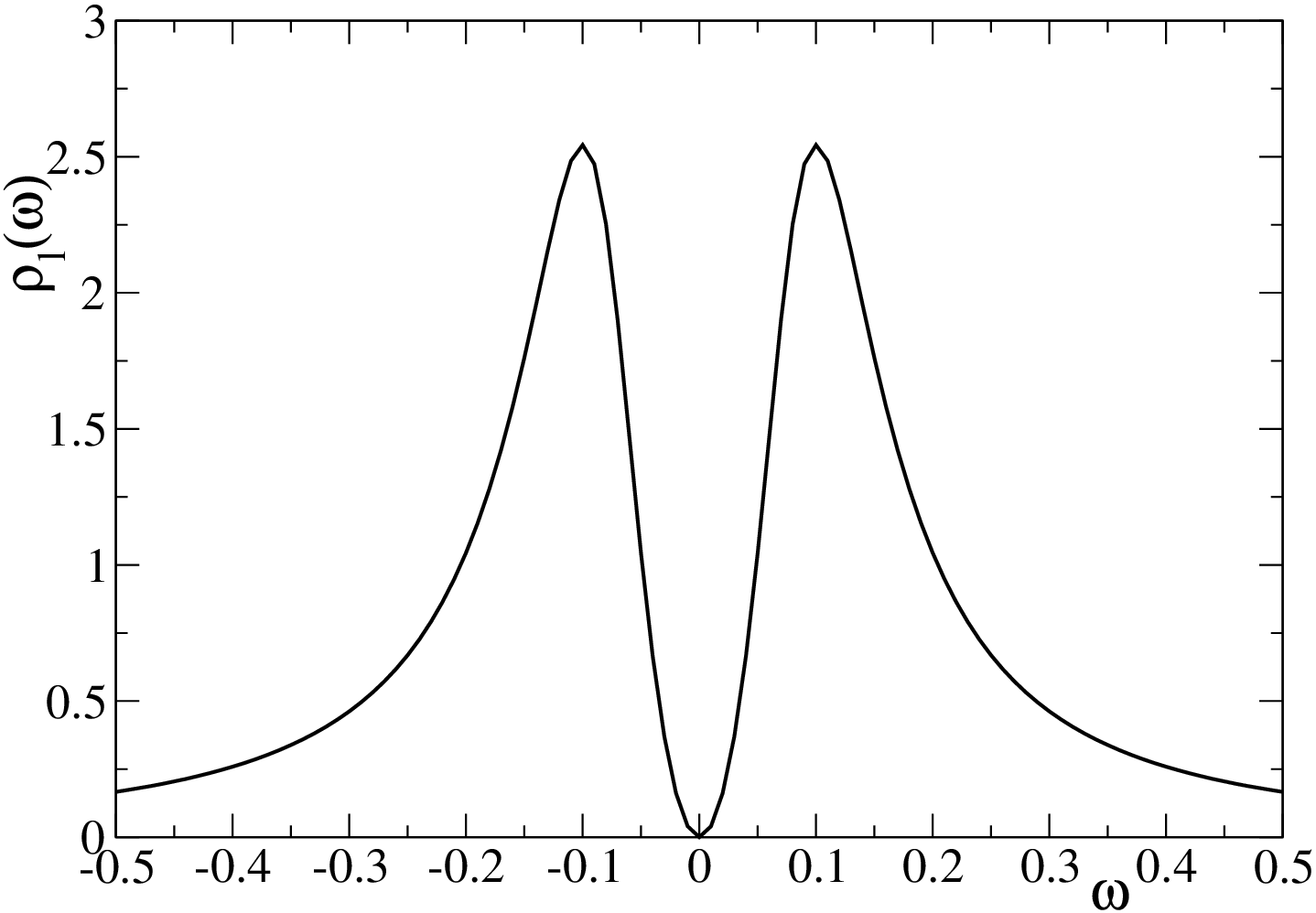}\\[0.5cm]
\includegraphics[width=0.8\columnwidth]{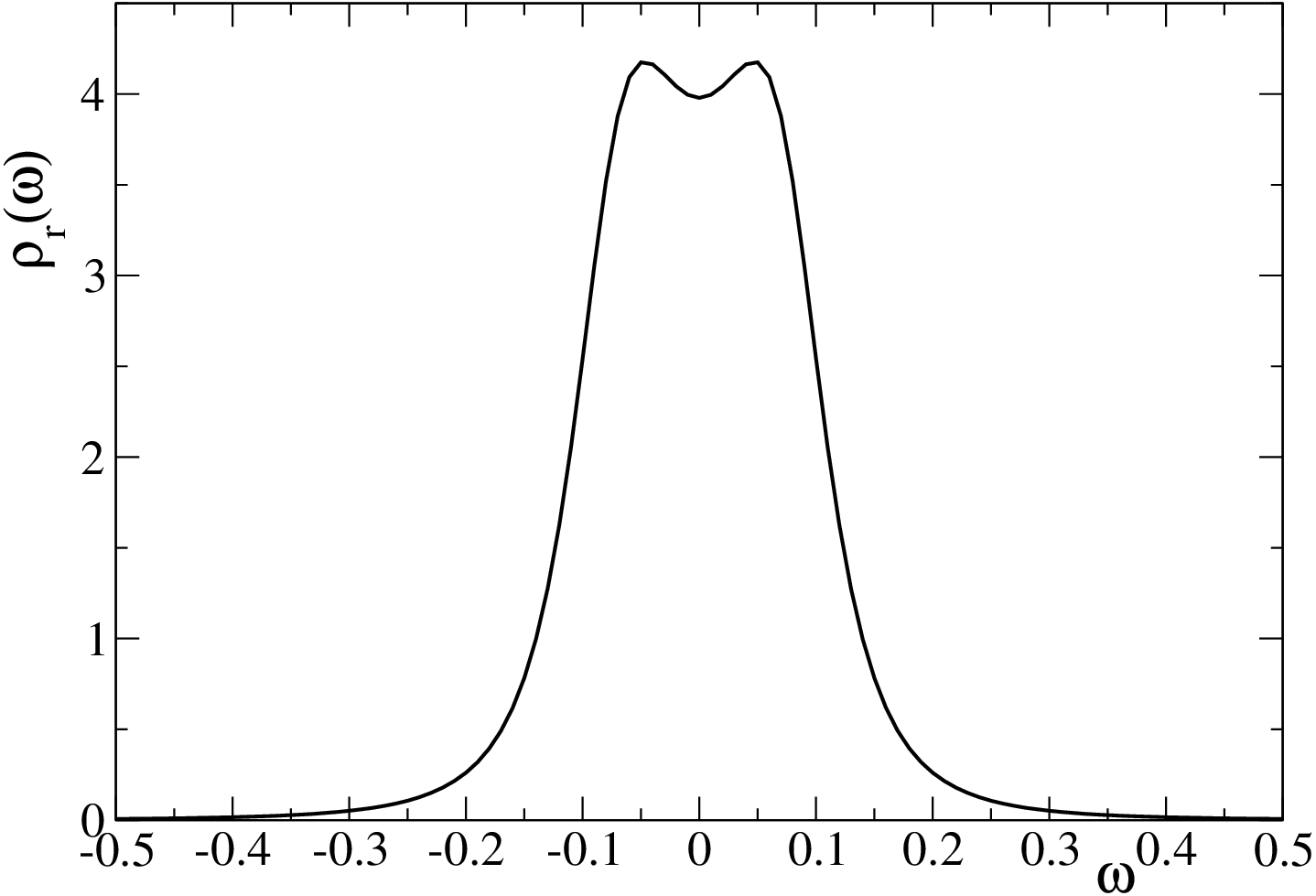}
\caption{Spectral density
of the left (top) and right (bottom) end state of the SSHC
connected at the left to a metallic lead. Parameters are $J=1$, $t_{lr}=0.1$ and $t_{cl}=0.5$.}
\label{rolr}
\end{figure}

The spectral densities for both end states are shown in Fig. \ref{rolr}. In contrast to the density at site 0 of the metallic lead, the density of the left end state vanishes at $\omega=0$
and is peaked at $\omega= \pm t_{lr}$. In comparison
the density of the right end mode is narrower and with a rather
flat maximum in the interval $-t_{lr}/2 < \omega < t_{lr}/2$.

\vspace{0.4cm}

\section{Summary and discussion}
\label{sum}

We have analyzed the effect of the splitting of the end modes
of a finite SSHC connected perpendicularly to a metallic lead on the
spectral densities and conductance through the lead at low energies. We have followed an analytical approach valid for energies below the gap of the isolated SSHC including in
particular accurately the localized modes at the ends of the SSHC.
At higher energies, provided the bandwidth of the metallic lead is sufficiently large, hybridization between the extended SSHC states and the metallic-lead states at comparable energies modifies the results.

The zero-temperature conductance through the metallic lead exhibits sharp zeros when the chemical potential coincides with the energies of the split end modes. This feature could be exploited to determine these energies experimentally.

The spectral density of the end state farthest from the metallic lead exhibits a narrow peak centered at zero energy, with a relatively flat maximum extending over an energy range of the order of half the splitting. In contrast, the spectral density of the end state coupled to the lead is broader and vanishes at zero energy.

\section*{Acknowledgments}
We thank Dunkan Mart\'{\i}nez Camacho for useful comments.

\bibliography{ref}

\end{document}